\documentclass[11pt]{article}
\usepackage{amsmath,amssymb,graphicx,cite,color,url,hyperref,setspace,multirow,bm}
\usepackage{bold-extra}
\begin{document}
\begin{flushright}
TIFR/TH/18-9
\end{flushright}

\begin{center}

\makebox{{\bfseries{\scshape{\Large Pinning down Anomalous 
$\bm{WW\gamma}$ Couplings at the LHC}}}}

{\sl Disha Bhatia}$^1$, {\sl Ushoshi Maitra}$^2$ 
and {\sl Sreerup Raychaudhuri}$^3$ \\ [2mm]
\setstretch{1.05}
{\small Department of Theoretical Physics, Tata Institute of Fundamental 
Research,\\ 1 Homi Bhabha Road, Mumbai 400 005, India.} \\

\setstretch{1.2}
\today

\bigskip

\centerline{\Large\bf Abstract}
\end{center}
\vspace*{-0.3in}
\begin{quotation}
\small
\setstretch{1.05}
\noindent We make a careful analysis of $W^\pm\gamma$ production at the 
LHC, identifying the $W^\pm$ through leptonic decays, with a view to 
exploring the sensitivity of the machine to anomalous $CP$-conserving 
$WW\gamma$ interactions. All the available kinematic variables are used, 
but we find that the most useful one is the opening angle in the 
transverse plane between the decay products of the $W^\pm$. It is shown 
that even a simple-minded analysis using this variable can lead to a 
much greater sensitivity at the LHC than the current constraints on the 
relevant parameters.
\end{quotation}
\normalsize
\setstretch{1.2}

The initial euphoria over the Higgs boson discovery of 2012\cite{Higgs-discovery} 
has now more-or-less abated, and even after more than a year's running of the CERN Large 
Hadron Collider (LHC) at the upgraded energy of 13 TeV, there have been 
no signs of new physics beyond the Standard Model (SM)\cite{LHC-negative}. 
However, while 
it has become abundantly clear that the belief that new particles and 
interactions would be discovered as soon as the LHC upgrade began to run 
was overly optimistic, there is no reason for despondency --- as 
yet\cite{Hewett-Moriondsummary}. 
This is because most of the new physics models proposed are of the 
decoupling type, with (possibly) highly massive particles and very 
feeble interactions, and may therefore prove much more difficult to 
discover than we have hitherto imagined -- or hoped. At this juncture, 
we may quickly fortify ourselves by noting that the last serendipitous 
discovery of an elementary particle (the $\tau$-lepton) occurred more than 
forty years ago, and that both the Higgs boson and the phenomena of neutrino 
oscillations took about the 
same time or even longer to establish experimentally. It is probably 
necessary, therefore, for high energy physicists to settle down for a 
long, hard grind before the expected new physics effects can be 
observed. For exist they must, if our ideas about quantum field theory, 
gravitation and cosmology are at all correct\cite{new-physics}. In any case, 
that a theory with as many ad hoc features as the SM can be the ultimate 
truth about Nature is unacceptable to many.

If we assume that there {\it is} new physics, but 
it consists of particles too massive to be discovered at the LHC, at 
least in the early stages of its 13~TeV run, then the only way in which 
these particles can be observed is through quantum effects, either at 
the tree or the loop levels. These will appear as modifications to the 
SM vertices, or the appearance of new, often higher-dimensional, 
operators involving the SM fields, with coefficients which are rendered 
small by the heavy mass scale of the underlying physics. Such {\it 
effective field theories} -- involving only the field content of the SM 
-- seem to offer the most promising window into physics beyond the SM\cite{EFT}. 
However, effective theories have their own problem. Most so-called 
UV-complete models beyond the Standard Model have only a limited  set of 
operators because of the twin constraints of gauge invariance (or 
extended gauge symmetries)

\vfill
\hrule
\vspace*{-0.1in}
{\small E-mail: $^1${\sf disha@theory.tifr.res.in} \hspace{0.65in}
{\sf $^2$ushoshimaitra32@gmail.com} \hspace{0.65in}
{\sf $^3$sreerup@theory.tifr.res.in}} \\

\newpage

and renormalisability. In contrast, effective field theories may
have a low-lying cutoff, which removes the requirement of 
renormalisability and permits a proliferation of operators -- all with 
small, but unknown coefficients. With so many unknown parameters, and 
only a finite set of measurables, almost any phenomenon can, in general, 
be explained and almost any prediction can be made. This is, if 
anything, a worse situation than the minimal SM even with all its ad hoc 
features.

It follows from the above that the broad picture of effective field 
theories is not perhaps the best approach to probe of physics beyond the SM. 
The focus in recent times has been, therefore, on a more minute 
examination of the operators, and on measurables which depend 
significantly on only a limited set of these operators, rather than the 
whole set -- an exercise which goes under the misnomer of {\it 
simplified models}, for it is the examination rather than the model 
which is simplified. Perhaps one of the earliest of these focussed 
examinations has been that of {\it anomalous} triple gauge-boson 
couplings (TGCs) \cite{Hagiwara,Hagiwara2}, which started from the days of 
the LEP collider\cite{Baur} and have acquired new relevance in the present 
climate\cite{Monalisa,Rosca,Hcorr}. These are anomalous, 
of course, only in the sense of being absent in the SM at tree level. 
The TGC's which have been considered are possible modifications 
to the $W^+W^-\gamma$ and $W^+W^-Z$ vertices, and possible new $ZZ\gamma$, 
$Z\gamma\gamma$ and $ZZZ$ vertices \cite{review}.

\begin{figure}[!ht]
\begin{minipage}{0.4\textwidth}
\begin{center}
\includegraphics[width=0.9\textwidth]{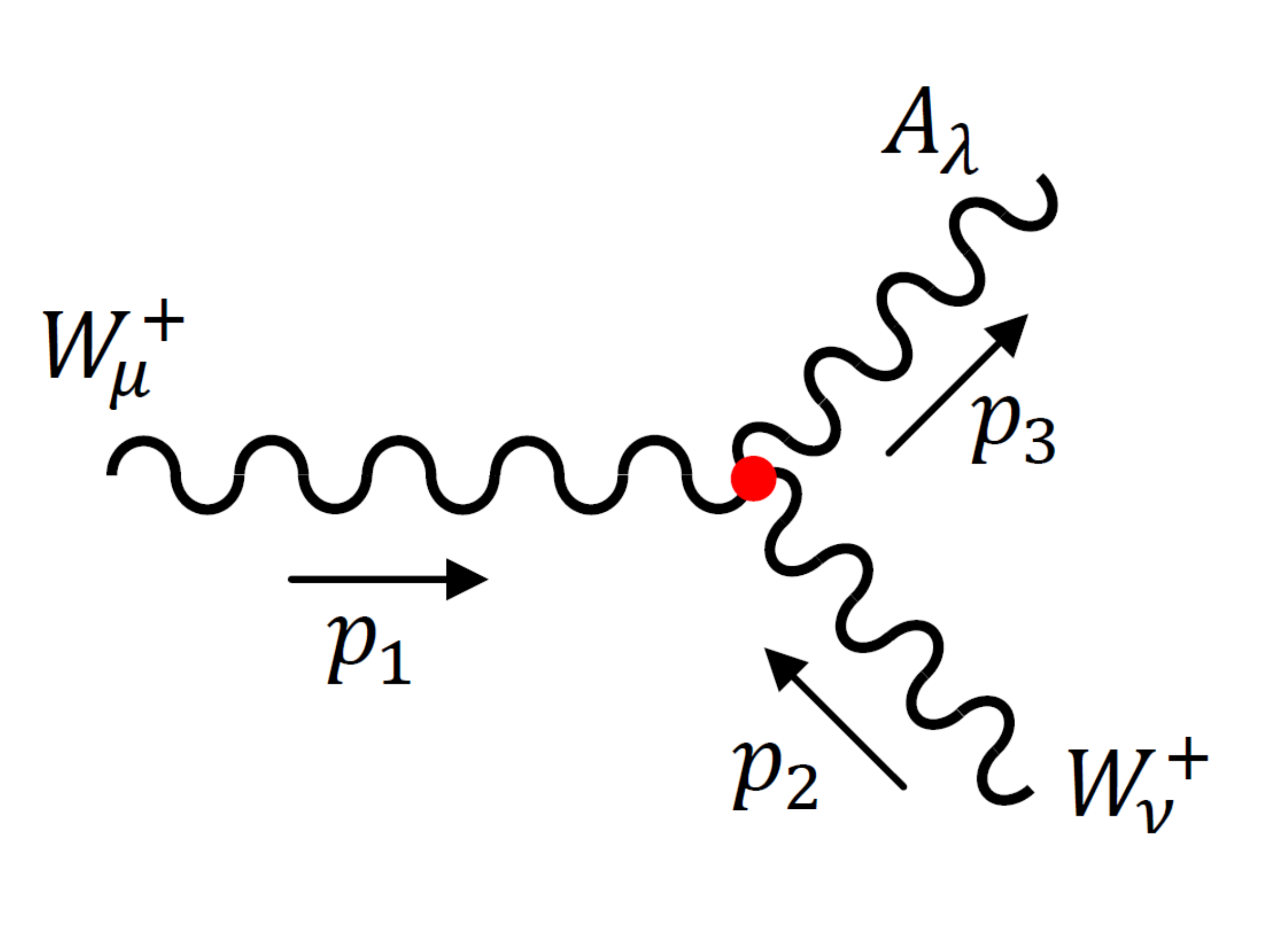}
\end{center}
\end{minipage}
\hskip 0.01\textwidth
\begin{minipage}{0.6\textwidth}
\setstretch{1.15}
This article takes just one of these vertices, viz, the 
$W^+W^-\gamma$ vertex illustrated on the left and considers a specific 
final state which is affected by only changes to this vertex. The 
process in question is
$$
p + p \to W^\pm + \gamma
$$ 
where the $W^\pm \to \ell^\pm \nu_\ell (\bar{\nu}_\ell)$ for $\ell = 
e,\mu$ and perhaps $\tau$.
\end{minipage}
\end{figure} 
If we denote the $W^+_\mu W^-_\nu A_\lambda$ vertex by 
$i\Gamma^{(WW\gamma)}_{\mu\nu\lambda}$, then the most general 
$CP$-conserving form consistent with the gauge and Lorentz symmetries of 
the SM can be parametrised\cite{Hagiwara} in the form of three separate 
terms, viz.
\begin{equation}
i\Gamma^{WW\gamma}_{\mu\nu\lambda}(p_1,p_2,p_3) 
= ie\left[T^{(0)}_{\mu\nu\lambda}(p_1,p_2,p_3)
+ \Delta\kappa_\gamma T^{(1)}_{\mu\nu\lambda}(p_1,p_2,p_3) 
+ \frac{\lambda_\gamma}{M_W^2} T^{(2)}_{\mu\nu\lambda}(p_1,p_2,p_3) 
\right]
\label{eqn:WWAvertex}
\end{equation}
where the $T_{\mu\nu\lambda}$ tensors are, respectively,
\begin{eqnarray}
T^{(0)}_{\mu\nu\lambda} &=& 
  g_{\mu\nu}  \left( p_1 - p_2  \right)_\lambda 
+ g_{\nu\lambda} \left( p_2 - p_3  \right)_\mu  
+ g_{\lambda\mu} \left( p_3 - p_1  \right)_\nu  
 \\
T^{(1)}_{\mu\nu\lambda} &=& 
g_{\lambda\mu} p_{3\nu} - g_{\nu\lambda} p_{3\mu}  
\nonumber \\
T^{(2)}_{\mu\nu\lambda} &=& 
p_{1\lambda}p_{2\mu}p_{3\nu} - p_{1\nu}p_{2\lambda}p_{3\mu}
- g_{\mu\nu} \left(p_2\cdot p_3 \ p_{1\lambda} - p_3\cdot p_1 \ p_{2\lambda}\right) 
\nonumber \\
& - & g_{\nu\lambda} \left(p_3\cdot p_1 \ p_{2\mu} - p_1\cdot p_2 \ p_{3\mu}\right) 
- g_{\mu\lambda} \left(p_1\cdot p_2 \ p_{3\nu} - p_2\cdot p_3 \ p_{1\nu}\right) 
\nonumber
\label{eqn:TGCs}
\end{eqnarray}
The tensor $T^{(0)}_{\mu\nu\lambda}$ in Eq.~(\ref{eqn:TGCs}) corresponds 
to the Standard Model coupling, while the tensors 
$T^{(1)}_{\mu\nu\lambda}$ and $T^{(2)}_{\mu\nu\lambda}$ give rise to 
{\it anomalous} TGC's. It may be noted that the dimension-4 tensor 
$T^{(1)}_{\mu\nu\lambda}$ can be absorbed in $T^{(0)}_{\mu\nu\lambda}$ 
with a coefficient $\kappa_\gamma = 1 + \Delta\kappa_\gamma$. However, 
in our work we have kept these tensors distinct as representing the SM 
and beyond-SM parts. Thus $\Delta\kappa_\gamma$ and $\lambda_\gamma$ 
parametrise the strength of these beyond-SM contributions --- which agrees with the common 
usage by most experimental collaborations\footnote{Strictly speaking, there
are SM contributions to $\Delta\kappa_\gamma$ and $\lambda_\gamma$ at 
higher orders. For example, at the one-loop level, there could be 
contributions of the order of (few)$\times 10^{-4}$ at a centre-of-mass
energy of TeV strength\cite{LEP-Kneur}. These are negligible in the current experimental 
studies, which, till date, only put constraints at the level of $10^{-2}$.}. 
It is reasonable to assume 
that $\Delta\kappa_\gamma$ will not be more than a few percent, for 
otherwise these corrections would have been detected when the $W$ itself 
was discovered, or when its properties were precisely measured at the CERN 
LEP-2\cite{LEP-2} and the Fermilab Tevatron\cite{Tevatron}. It is also 
traditional to parametrise the 
mass-suppression of the dimension-6 operator $T^{(2)}_{\mu\nu\lambda}$ 
with a factor $M_W^{-2}$. However, if the operator arises from new physics at a 
scale $\Lambda$, the corresponding coefficient should have been 
$\xi/\Lambda^2$, where $\xi$ is some coupling -- perhaps ${\cal O}(1)$ --
and hence, we can identify
\begin{equation}
\lambda_\gamma = \xi \left( \frac{M_W}{\Lambda} \right)^2
\end{equation}
In fact, setting $\xi = 1$, and $\Lambda = 1$~TeV, we get $\lambda_\gamma \simeq 
0.0065$. We may thus expect $\lambda_\gamma$ to lie an order of 
magnitude below $\Delta \kappa_\gamma$, and, in fact, we shall see
below that this is indeed true for the experimental constraints. 

We remark in passing that there are also $CP$-violating 
contributions to the $W^+W^-\gamma$ vertex, which can be parametrised in 
terms of two coupling constants $\tilde{\kappa}_\gamma$ and 
$\tilde{\lambda}_\gamma$. However, these are already constrained to be very 
small from the measurement of the electric dipole moment of the 
neutron\cite{CPV-TGC}, and hence we will not consider them further in this article. 
It is also possible -- in fact, plausible -- that if the photon has 
anomalous couplings with a $W^+W^-$ pair, then the $Z$ boson may also 
have such anomalous couplings, which may be related in some way by the 
gauge invariance of the SM\cite{Hagiwara}. However, the philosophy adopted in this 
article is that these will not affect the measurement in question, and 
can therefore, be kept outside the scope of the discussion. Experimental
bounds involving $WW$ production \cite{LEP-2,ATLAS-WW,CMS-WW} 
have to consider this possibility and hence
always carry a caveat about the choice of $WWZ$ couplings. 

\vspace*{-0.2in} 
\begin{center} 
\begin{figure}[htb] 
\centerline{\includegraphics[width=0.65\textwidth]{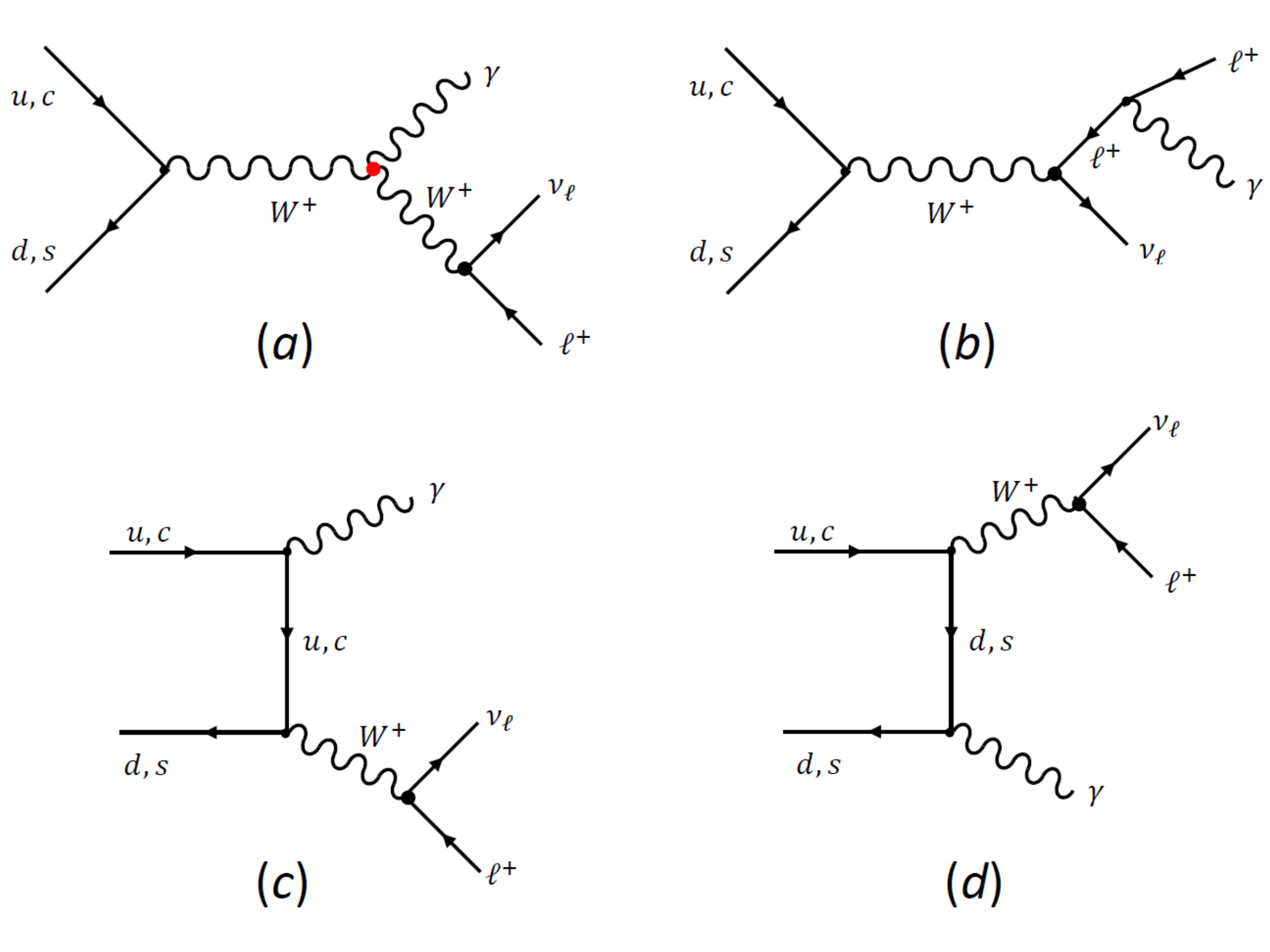} } 
\vspace*{-0.1in} 
\caption{\setstretch{1.05}\footnotesize Feynman 
diagrams contributing to the final state $\gamma\ell^+\nu_\ell$ at a 
hadron collider, with initial $u\bar{d}$ (or the more suppressed 
$c\bar{s}$) partonic states. These diagrams correspond to both the 
signal and the background, since the $W^+W^-\gamma$ vertex, indicated by 
the red dot in diagram (a), has both SM and anomalous contributions. } 
\label{fig:diagrams} 
\end{figure} 
\end{center} 
\vspace*{-0.5in}

The production of a $W^\pm$ associated with a hard transverse photon is 
one of the most standard processes which one considers at a hadron 
collider\cite{CMS-WA,ATLAS-WA}. It occurs through a pair of dissimilar quarks, 
e.g. $u$ and 
$d$, as the initial-state partons, which are required for single $W$ 
boson production. A photon can then be radiated off any of the internal 
or external legs of the corresponding diagram. However, if we allow the 
$W^\pm$ to decay further into a charged lepton $\ell^\pm$ and the 
corresponding neutrino $\nu_\ell$, we will have one more diagram where a 
photon is radiated off the charged lepton. The final state consists, 
then, of a photon, a lepton and missing energy from the neutrino. One 
would also require a jet veto to keep the process hadronically quiet. 
The four diagrams at leading order are illustrated in Fig.~\ref{fig:diagrams}.  
 
When these diagrams are evaluated, the Feynman amplitude can be written 
\begin{equation} 
{\cal M}(\Delta\kappa_\gamma, 
\lambda_\gamma) = {\cal M}_{\rm SM} + \Delta\kappa_\gamma {\cal 
M}_\kappa + \lambda_\gamma {\cal M}_\lambda 
\label{eqn:matrix} 
\end{equation} 
squaring which, it follows that the cross-section will be 
a combination of terms 
\begin{eqnarray} 
\sigma(\Delta\kappa_\gamma, \lambda_\gamma) & = & \sigma_{\rm SM} + 
\left(\Delta\kappa_\gamma\right)^2 \sigma_\kappa + \lambda_\gamma^2 
\sigma_\lambda \nonumber \\ 
&& + \ \Delta\kappa_\gamma 
\sigma_{\kappa,{\rm SM}} + \lambda_\gamma \sigma_{\lambda,{\rm SM}} + 
\Delta\kappa_\gamma \lambda_\gamma \sigma_{\kappa,\lambda} 
\label{eqn:csecn} 
\end{eqnarray} 
where the terms on the first line of 
Eq.~(\ref{eqn:csecn}) arise from the squares of the corresponding terms 
in Eq.~(\ref{eqn:matrix}), while the terms on the second line are the 
respective interference terms. Since $\Delta\kappa_\gamma$ and 
$\lambda_\gamma$ are small, it is clear that $\sigma_{\rm SM}$ will be 
the dominant term -- or dominant background -- while the other terms in 
Eq.~(\ref{eqn:csecn}) will constitute a small signal. Of these, the 
terms linear in $\Delta\kappa_\gamma$ and $\lambda_\gamma$ will 
generally be the largest. The challenge is, therefore, to isolate the 
extremely small signal from the large SM background by the judicious use 
of kinematic cuts and distributions. At this point, we note that QCD
corrections to the $W\gamma$ process may increase\cite{QCDcorrection} the overall 
cross-section by $30 - 40$\%. However, these may be expected to be rather 
similar for both signal and background, and hence are not taken into account 
in our analysis.

In the experimental situation, our concern is with a hadronically-quiet 
final state consisting of a hard transverse photon, a hard transverse 
lepton and substantial missing energy. This is a very clean signal and, 
barring issues like pileup and multiple interactions at the LHC, may be 
expected to constitute a strong probe for the underlying physics -- in 
this case, the TGC concerned. Since the final state is so simple, there 
exists only a small number of kinematic variables which are invariant under 
longitudinal boosts, and these, together with the cuts we have imposed 
on them, are listed below.
\vspace*{-0.2in}
\begin{enumerate}

\item[(A)] The magnitude of the transverse momentum of the photon 
($p_{T\gamma}$), which we require to satisfy $p_{T\gamma} \geq 60$~GeV.
\item[(B)] The pseudorapidity of the photon ($\eta_\gamma$), which we 
require to satisfy $\eta_\gamma \leq 2.5$.
\item[(C)] The magnitude of the transverse momentum of the lepton 
($p_{T\ell}$), which we require to satisfy $p_{T\ell} \geq 40$~GeV.
\item[(D)] The pseudorapidity of the lepton ($\eta_\ell$), which we 
require to satisfy $\eta_\ell \leq 2.5$.
\item[(E)] The magnitude of the missing transverse momentum 
($\not{\!p}_T$), which we require to satisfy $\not{\!p}_T \geq 40$~GeV.
\item[(F)] The so-called angular separation between photon and lepton 
($\Delta R_{\gamma\ell}$), which we require to satisfy $\Delta 
R_{\gamma\ell} \geq 0.4$.
\end{enumerate}
\vspace*{-0.2in}
The cuts in (A) -- (E) are driven more by ease of identification of the 
final state and the detector coverage, while (F) is included to suppress 
the collinear photons which are preferred by the SM diagrams in 
Fig~\ref{fig:diagrams}(b)-(d).

In addition to the above, if we consider the vector momenta in the 
transverse plane, which we denote $\vec{p}_{T\gamma}$, $\vec{p}_{T\ell}$ 
and $\vec{\not{\!p}}_T$, we can construct three more variables which are 
invariant under longitudinal boosts. These are
\begin{equation}
\Delta \varphi_{\gamma\ell} = \cos^{-1}\left(
\frac{\vec{p}_{T\gamma} \cdot \vec{p}_{T\ell} }{p_{T\gamma} \ p_{T\ell} }\right)
\qquad
\Delta \varphi_{\gamma\not{p}_T} = \cos^{-1}\left(
\frac{\vec{p}_{T\gamma} \cdot \vec{\not{\!p}}_T }{p_{T\gamma} \ \not{\!p}_T }\right)
\qquad
\Delta \varphi_{\ell\not{p}_T} = \cos^{-1}\left(
\frac{\vec{p}_{T\ell} \cdot \vec{\not{\!p}}_T }{p_{T\ell} \ \not{\!p}_T }\right)
\label{eqn:azimuth}
\end{equation}
These angular variables are known to be highly sensitive to 
momentum-dependent operators\cite{Monalisa} and since the tensors 
$T^{(1,2)}_{\mu\nu\lambda}$ are of this kind, we may expect them to 
carry some signs of the anomalous TGCs. In fact, we find that the only 
variables which are sensitive to these are the transverse momenta in 
(A), (C) and (E) above, and these angular variables in 
Eq.~(\ref{eqn:azimuth}).

\footnotesize
\begin{table}[!h]
\begin{center}
\begin{tabular}{crrr}
Cut   & $\sigma_{\rm SM}$ & $\sigma_\kappa$ & $\sigma_\lambda$  \\ [1mm]
\hline\hline
$p_{T\gamma} \geq 60$~GeV &  430.11~fb  &   737.82~fb  &  41.89~pb \\ [1mm]
\hline
$p_{T\gamma} \geq 60$~GeV &  100.0\;\%  &    100.0\;\%  &    100.0\;\%  \\ [1mm]
$p_{T\ell} \geq 40$~GeV  & 70.25\;\%   &   75.70\;\%   &   85.55\;\% \\ [1mm]
$\not{\!p}_T \geq 40$~GeV & 22.82\;\%   &   52.34\;\%   &   70.77\;\% \\ [1mm]
$M_{TW} \geq 30$~GeV & 20.68\;\%   &   43.13\;\%   &   55.11\;\% \\ [1mm]
$\eta_\gamma \leq 2.5$ & 15.89\;\%   &   36.88\;\%   &   53.50\;\% \\ [1mm]
$\eta_\ell \leq 2.5$ & 12.28\;\%   &   32.61\;\%   &   52.24\;\% \\  [1mm]
$\Delta R_{\gamma\ell} \geq 0.4$ & 11.30\;\%  &   32.60\;\%   &   52.26\;\% \\ [1mm]
\hline
    & 48.57~fb  &  240.52~fb  &  21.89~pb \\
\hline 
\end{tabular}
\caption{\footnotesize Cut flow table showing the effect of different 
kinematic cuts on the principal terms in the cross-section. As may be 
expected, the SM contribution is brought down to about one tenth, 
whereas the others are reduced to roughly a third and a fifth 
respectively. The large value of $\sigma_\lambda$ is due to the 
inappropriate choice of $M_W^2$ as the suppression factor --- if we had 
chosen $\Lambda = 1$~TeV instead, these cross-sections would be 
suppressed by a factor $\left(M_W/{\rm 1~TeV}\right)^2 \approx 6.4\times 
10^{-3}$, which would bring them on par with the previous columns.}
\label{tab:cutflow}
\end{center}
\end{table}
\vspace*{-0.2in}
\normalsize
Finally, to ensure good convergence of our Monte Carlo simulations, we 
construct\cite{ATLAS-WA,CMS-WA} the variable $M_{TW}$, where
\begin{equation}
M_{TW}^2 = 2 \ p_{T\ell} \not{\!p}_T \left( 1 - \cos \Delta 
\varphi_{\ell\not{p}_T} \right)
\end{equation}
and impose a cut $M_{TW} \geq 30$~GeV. 
The effect of these successive kinematic cuts on the terms in the cross-section 
is shown in Table~\ref{tab:cutflow}.
Any stronger cuts would result in severe loss in the TGC signal, both 
for $\Delta\kappa_\gamma$ and $\lambda_\gamma$.
\vspace*{-0.2in}
\begin{center}
\begin{figure}[!htb]
\centerline{\includegraphics[width=0.96\textwidth,height=0.48\textheight]{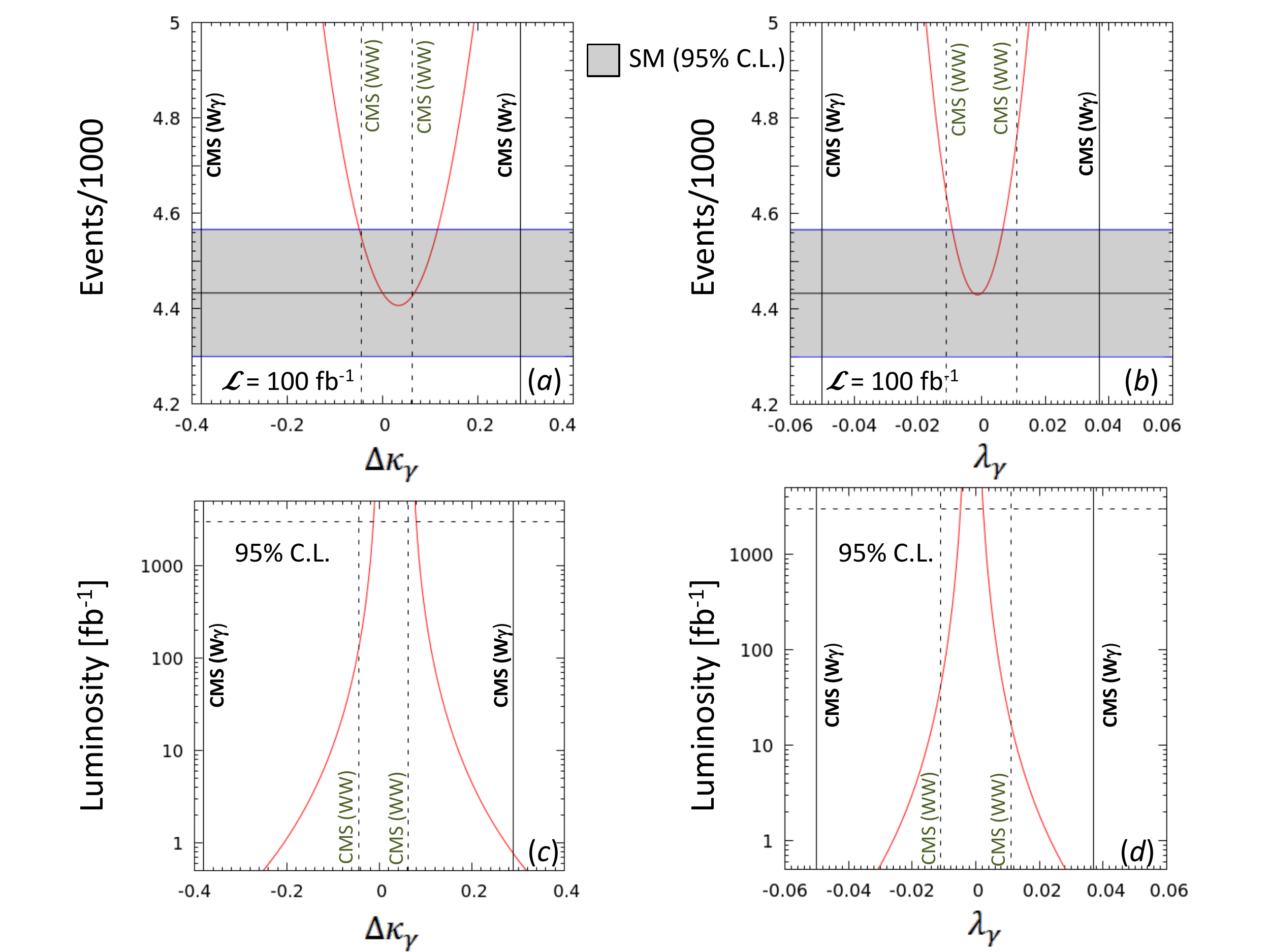} }
\vspace*{-0.1in}
\caption{\setstretch{1.05}\footnotesize Constraints on the anomalous
$WW\gamma$ couplings from consideration of the total cross-section,
assuming an integrated luminosity of 100~fb$^{-1}$. The upper panels 
correspond to the variation in the excess in events per thousand over 
the SM prediction for the cases 
(a) $\Delta\kappa_\gamma\neq 0, \lambda_\gamma = 0$ and 
(b) $\Delta\kappa_\gamma = 0, \lambda_\gamma \neq 0$. The horizontal 
line shows the SM prediction and the shaded portion corresponds to its
variation at 95\% C.L. Solid vertical lines marked `CMS(W$\gamma$)' 
correspond to the Run-1 CMS bounds on the corresponding anomalous 
coupling from $W\gamma$ production\cite{CMS-WA} and broken verticals 
marked `CMS(WW)' 
correspond to similar bounds obtained from $WW$ production\cite{CMS-WW}, 
assuming that
$WW\gamma$ and $WWZ$ anomalous couplings are related through $SU(2)$
symmetry. The lower panels, marked (c) and (d) respectively, show the 
corresponding 95\% C.L. search limits (see text) when the luminosity 
is varied up to 5~ab$^{-1}$, with a horizontal broken line to indicate
the machine limit of 3~ab$^{-1}$ for the HL-LHC.}
\label{fig:total-cs}
\end{figure}
\end{center}
\vspace*{-0.4in}

If we consider the total cross-section, as given above, the limits one can put 
on the parameters $\Delta\kappa_\gamma$ and $\lambda_\gamma$ are already 
strong. The actual number of signal events (in thousands) expected are 
shown in the panels marked ($a$) and ($b$) in 
Fig.~\ref{fig:total-cs}, assuming an integrated luminosity of 100 
fb$^{-1}$. The abscissa in ($a$) and ($b$) shows, respectively, the values 
of $\Delta \kappa_\gamma$ and $\lambda_\gamma$, each assuming that the other is 
zero. The region marked in grey corresponds to the 95\% confidence
level (C.L.) fluctuation in the SM prediction. Solid vertical lines 
indicate the current experimental bounds from $W\gamma$ production
at the LHC\cite{CMS-WA,ATLAS-WA}, which directly constrains the 
$WW\gamma$ vertex, whereas broken vertical lines indicate the bounds 
from $WW$ production\cite{ATLAS-WW,CMS-WW},
where there are contributions from both $WW\gamma$ and $WWZ$ vertices.
As explained above, these constraints are not as solid as those
obtained from $W\gamma$ production. However, it is immediately
obvious that the signal considered in this work can achieve the 
95\% C.L. level even at values which are comparable with the $WW$
constraints, and certainly far smaller than the current $W\gamma$
constraints.
     
If the plots in the upper panels of  Figure~\ref{fig:total-cs} indicate strong 
constraints with a luminosity of 100~fb$^{-1}$, it is relevant to ask what may
be achieved at the high-luminosity upgrade of the LHC (HL-LHC), where 
the integrated luminosity may go as high as 3~ab$^{-1}$. To determine
the search limits, we can determine the signal significance 
$\chi^2(L,\Delta\kappa_\gamma, \lambda_\gamma)$ as a function of 
luminosity $L$ as
\begin{equation}
\chi^2(L,\Delta\kappa_\gamma, \lambda_\gamma) 
= \left[\frac{L\{\sigma(\Delta\kappa_\gamma, \lambda_\gamma) - \sigma_{SM}\}} 
{\sqrt{L\sigma_{SM}}}\right]^2
\label{eqn:chi2tot}
\end{equation}
assuming Gaussian random fluctuations in the background
$\delta (L\sigma_{SM}) = \sqrt{L\sigma_{SM}}$. For this study, we 
ignore systematic errors, or, more properly, assume that they will
be small enough to be ignored, compared to the statistical error.
Now if, for a given value of $L$, the value(s) of $\Delta\kappa_\gamma$ 
and/or $\lambda_\gamma$ satisfies $\chi^2(L,\Delta\kappa_\gamma, 
\lambda_\gamma) > 1.96$, we qualify the signal for an anomalous TGC as 
observable at 95\% C.L. The corresponding variations, for the cases
(c) $\Delta\kappa_\gamma\neq 0, \lambda_\gamma = 0$ and 
(d) $\Delta\kappa_\gamma = 0, \lambda_\gamma \neq 0$ are plotted
in the lower panels of Figure~\ref{fig:total-cs}. It may immediately
be seen that even with a very low integrated luminosity, the 13~TeV
LHC does immensely better than the Run-1 data, and with an integrated 
luminosity of 1~ab$^{-1}$, the direct constraints from the total
cross-section are better than those even from $WW$ production (which
involve the $WWZ$ couplings), except for one case 
$\Delta\kappa_\gamma > 0, \lambda_\gamma = 0$. At this juncture it
is relevant to note the asymmetry of the curves in each panel
about the zero point, which can be attributed to large interference
terms between the anomalous $WW\gamma$ operators and the SM ones.

We now address the principal question for which this work was taken
up, and that is whether the study of differential cross-sections
instead of the total cross-section can help better in identifying 
anomalous $WW\gamma$ couplings. We have made a careful study of 
practically all the straightforward kinematic variables it is
possible to construct with a $\gamma\ell\not{\!p}_T$ final state.
It turns out that the ones which are sensitive to the anomalous
couplings, i.e. the ones for which the anomalous operators behave
differently than the SM operators, are those listed below:
\begin{table}[!h]
\begin{center}
\begin{tabular}{c|c|c|c|c|c|c}
$X = $ &($a$)           & ($b$)         & ($c$)           & 
($d$)           & ($e$) & ($f$) \\ \hline
$v_X=$ &$p_{T\gamma}$ & $p_{T\ell}$ & $\not{\!p}_T$ &
$\Delta \varphi_{\gamma\ell}$ & $\Delta \varphi_{\gamma\not{\!p}_T}$ &
$\Delta \varphi_{\ell\not{\!p}_T}$ \\
\hline
\end{tabular}
\caption{\footnotesize List of kinematic variables whose distributions
are sensitive to anomalous TGCs.}
\label{tab:vars}
\end{center}
\end{table}
\vspace*{-0.4in} 

\begin{center}
\begin{figure}[!htb]
\centerline{\includegraphics[width=0.9\textwidth,height=0.45\textheight]{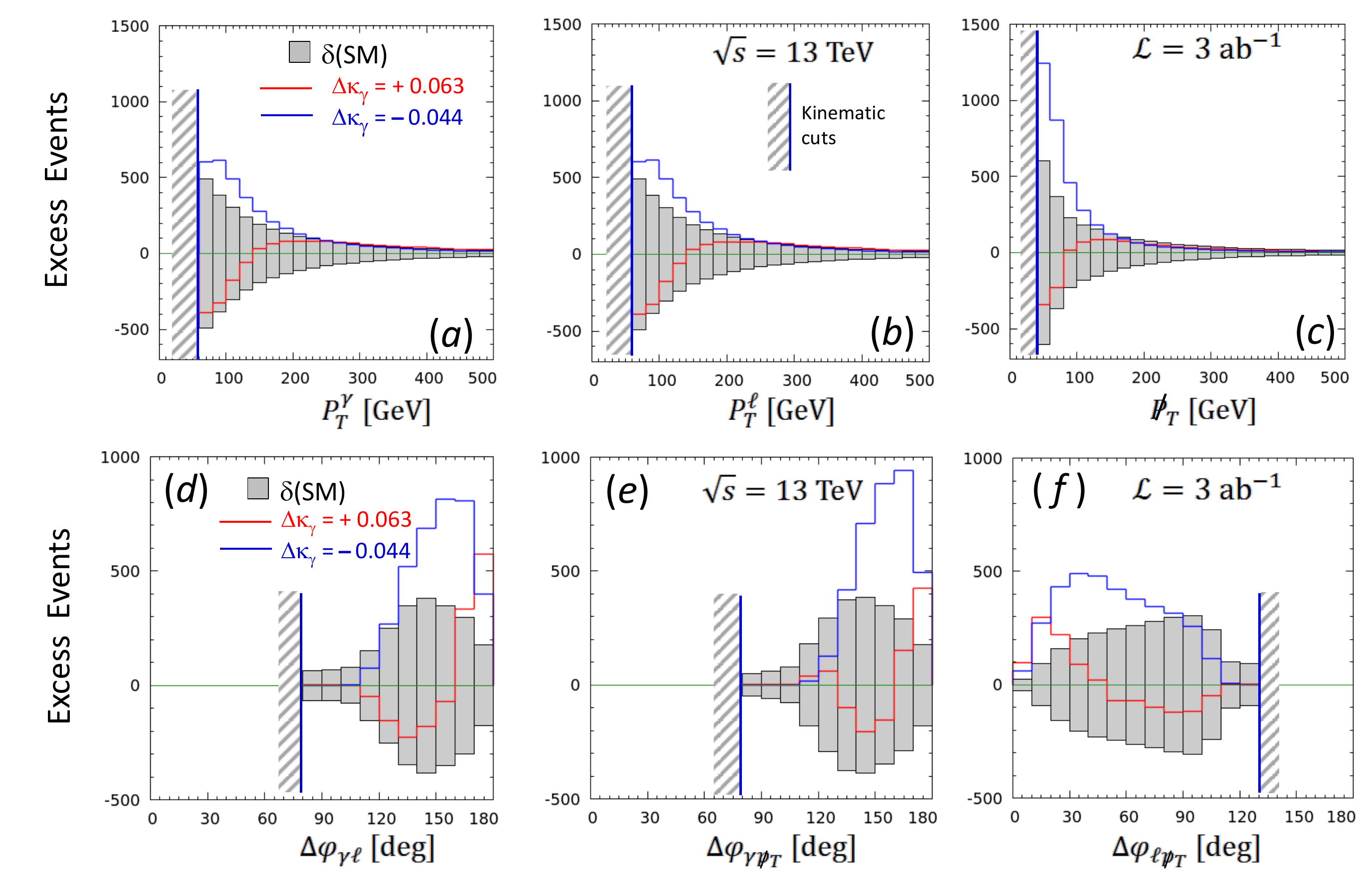} }
\caption{\setstretch{1.05}\footnotesize Background-subtracted 
kinematic distributions for the different variables listed above in 
the case of $\Delta\kappa_\gamma \neq 0$ with $\lambda_\gamma = 0$. 
The panels are marked (a), (b), etc. according to the legend in the 
text. Red histograms correspond to the signal with a positive value 
(marked) of $\Delta\kappa_\gamma$ and blue histograms correspond to 
negative values of $\Delta\kappa_\gamma$, while the shaded histograms
correspond to the 95\% C.L. fluctuations in the SM background. Vertical 
lines with hatching indicate the kinematic cuts listed in the text. }
\label{fig:distrib-kap}
\end{figure}
\end{center}

\begin{center}
\begin{figure}[!htb]
\centerline{\includegraphics[width=0.9\textwidth,height=0.45\textheight]{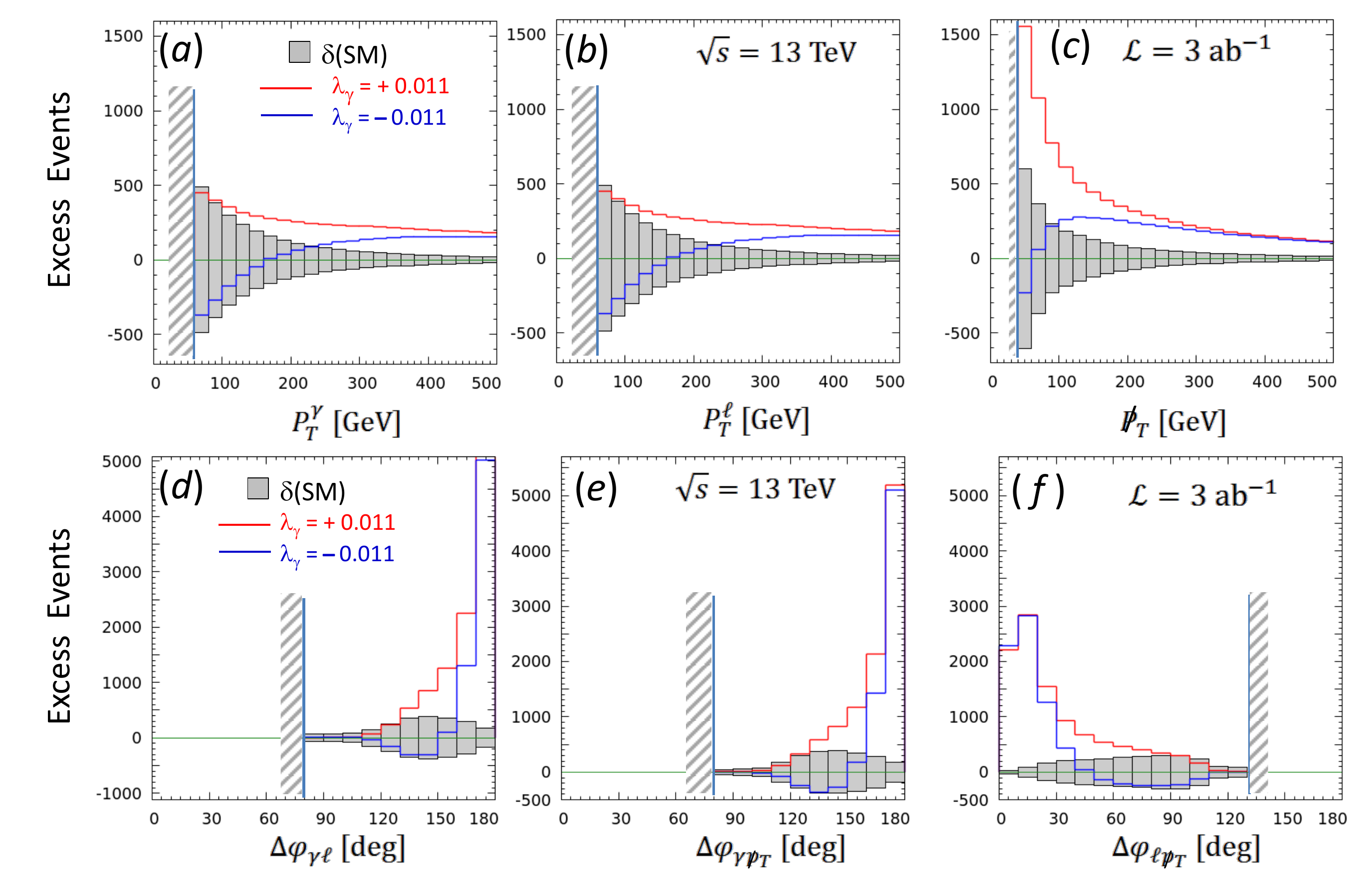} }
\caption{\setstretch{1.05}\footnotesize Background-subtracted 
kinematic distributions for the different variables listed above in 
the case of $\lambda_\gamma \neq 0$ with $\Delta\kappa_\gamma  = 0$. 
The panels are marked (a), (b), etc. according to the legend in the 
text. Red histograms correspond to the signal with a positive value 
(marked) of $\lambda_\gamma$ and blue histograms correspond to 
negative values of $\lambda_\gamma$, while the shaded histograms
correspond to the 95\% C.L. fluctuations in the SM background. Vertical 
lines with hatching indicate the kinematic cuts listed in the text.}
\label{fig:distrib-lam}
\end{figure}
\end{center}

The effect of the anomalous TGCs on these is, of course, different
for different observables, and this is illustrated in Figures~
\ref{fig:distrib-kap} and \ref{fig:distrib-lam}. In Figure~\ref{fig:distrib-kap} 
we show three histograms in each panel, for the bin-wise quantity 
\begin{equation}
N_{\rm excess} = L \left(\frac{d\sigma}{dv_X} 
- \frac{d\sigma_{\rm SM}}{dv_X} \right) \ ,
\label{eqn:distrib}
\end{equation}
where $L$ is the integrated luminosity and $v_X$ is the corresponding
variable in Table~\ref{tab:vars}. 
In each panel of Figure~\ref{fig:distrib-kap}, the red histogram 
corresponds to the excess events as per Eqn.~(\ref{eqn:distrib}) 
for $\Delta\kappa_\gamma = +0.063$, i.e.
the more stringent CMS upper limit arising from the $WW$ cross-section\cite{CMS-WW},
and the blue histogram indicates the corresponding lower limit
$\Delta\kappa_\gamma = -0.063$. The solid shaded region represents
the 95\% C.L. fluctuations in the SM prediction, denoted 
$\delta({\rm SM})$. In each case, the kinematic cuts listed in the 
text above are shown by a vertical line and hatching. For these plots,
we have set $L = 3$~ab$^{-1}$, i.e., the maximum envisaged value of the 
HL-LHC.          

If we consider the case of $\Delta\kappa_\gamma > 0$, i.e. the red 
histograms in Figure~\ref{fig:distrib-kap}, we can see that the
number of excess events is substantially above the SM fluctuation
for a significant number of bins, especially as one goes towards
higher magnitudes of $p_T$ and for back-to-back vectors in the transverse
plane, except for the opening angle in the transverse plane between
the decay products of the $W$, which tend to be aligned for the signal.
In fact, in some of the bins, the deviation is rather large. One the other 
hand, if we consider the case of $\Delta\kappa_\gamma < 0$, i.e. the blue 
histograms in Figure~\ref{fig:distrib-kap}, the deviations are large
only for really high magnitudes of $p_T$ and even more extreme angles in
the transverse plane than in the case of positive $\Delta\kappa_\gamma$.  

\vspace*{-0.5in}

Some of the salient features of the histograms in Figure~
\ref{fig:distrib-kap} are listed below.
\vspace*{-0.2in}
\begin{itemize}
\item In all the panels, the signal histograms for negative 
$\Delta\kappa_\gamma$ change sign over the selected range,
whereas for positive $\Delta\kappa_\gamma$ they are monotone. 
\item Of the upper three panels, clearly the best signal will
come from a study of the missing $p_T$, for, even for negative 
$\Delta\kappa_\gamma$, there are significant deviations
over 100~GeV.  
\item In the lower three panels, all show large deviations from
the SM background. It is not clear by inspection which of these three
variables is best suited to find the signal. For this, we must
develop a suitable numerical metric.
\end{itemize}
\vspace*{-0.1in}
We then turn to the other anomalous coupling $\lambda_\gamma$, in the 
case when $\Delta\kappa_\gamma = 0$. This is illustrated in Figure~
\ref{fig:distrib-lam}, where we show the same three histograms in each 
panel as for Figure~\ref{fig:distrib-kap}, for the bin-wise quantities as 
defined in Eqn.~(\ref{eqn:distrib}) and the table below it. The notations 
and conventions of Figure~\ref{fig:distrib-lam} are therefore identical 
with those of Figure~\ref{fig:distrib-kap}. Obviously the range of 
values of $\lambda_\gamma$ is smaller, but this is, as explained before,
due to the artificial scaling with $M_W$ instead of some higher scale.
Thus, the red (blue) histograms correspond to $\lambda_\gamma = 
+0.011(-0.011)$, which are, as before, the CMS limits from $WW$ production\cite{CMS-WW}.
Qualitatively, the deviations are rather similar to those in
Figure~\ref{fig:distrib-kap}, and one cannot tell, just by inspection,
which of the parameters is preferable. Thus, if indeed, a deviation
in these distributions from the SM prediction is found, we will encounter
a difficult {\it inverse problem}, i.e. separation of signals from
$\Delta\kappa_\gamma$ from those for $\lambda_\gamma$. In the present 
article, however, we feel that it is premature to address this problem. 
Instead, we focus on whether it will be possible to extend the discovery 
reach of the LHC by considering these distributions, rather than the total 
cross-section. The time to address this distinction will come when
a deviation is actually found.  

\vspace*{-0.5in}

In order to see if a distribution has enough deviation from the SM
prediction to be observable at, say, 95\% C.L., we need to construct
a suitable numerical metric. We choose a simple-minded extension of the one in 
Eqn.~(\ref{eqn:chi2tot}), in the form
\begin{equation}
\chi_X^2(L,\Delta\kappa_\gamma, \lambda_\gamma) = 
\sum_{n=1}^{N_X} \left(\frac{N^{(n)}_{\rm excess}}{\sqrt{N^{(n)}_{\rm SM}}}\right)^2
\label{eqn:chisqdistrib}
\end{equation}
where the index $n$ runs over all the bins, and 
\begin{equation}
N^{(n)}_{\rm SM} = L \frac{d\sigma^{(n)}_{\rm SM}}{dv_X} 
\end{equation}
is the SM prediction in that bin. $N_{\rm excess}$ is defined in
Eqn.~(\ref{eqn:distrib}), but here it carries a bin index $n$, 
and $L$ is, as usual, the integrated luminosity. The total number
of bins $N_X$ is not the same for all the different variables $v_X$,
as a glance at Figures~\ref{fig:distrib-kap} and \ref{fig:distrib-lam}
will show. We can now compare the calculated values of 
$\chi_X^2(L,\Delta\kappa_\gamma, \lambda_\gamma)$ with 
$\chi^2(N_X,95\%)$ which is the probability that the
SM cross-section with $N_X$ bins will undergo a 95\% Gaussian 
fluctuation, faking a signal. If, for a given set of arguments,
$\chi_X^2(L,\Delta\kappa_\gamma, \lambda_\gamma) > \chi^2(N_X,95\%)$,
we will assume the corresponding anomalous TGC to be discoverable
at the LHC.

\begin{center}
\begin{figure}[!htb]
\centerline{\includegraphics[width=0.85\textwidth,height=0.5\textheight]{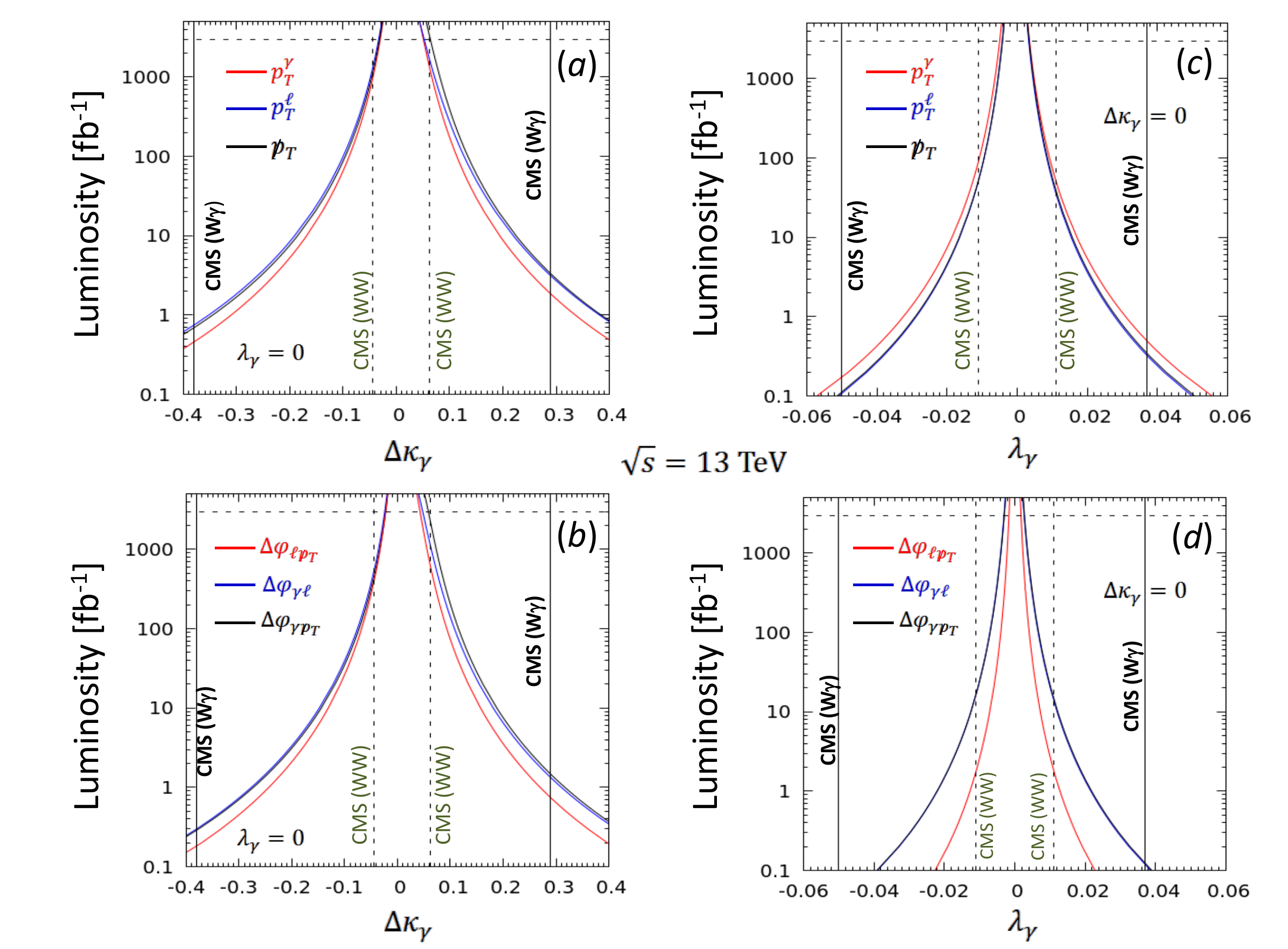} }
\vspace*{-0.1in}
\caption{\setstretch{1.05}\footnotesize 95\% C.L. discovery limits for the case
$\Delta\kappa_\gamma \neq 0, \lambda_\gamma = 0$ in the panels on the left, marked
($a$) and ($b$), and the case $\Delta\kappa_\gamma = 0, \lambda_\gamma \neq 0$ in the 
panels on the right, marked ($c$) and ($d$). Only three variables at a time have been
shown in each panel to avoid clutter. The upper panels, marked ($a$) and ($c$) show
the discovery limits for transverse momentum variables, while the lower panels,
marked ($b$) and ($d$) show the discovery limits for azimuthal angle variables.}
\label{fig:limits-single}
\end{figure}
\end{center}
\vspace*{-0.6in}

Our results for the different variables are shown in Figure~\ref{fig:limits-single}.
The upper panels, marked ($a$) and ($c$) show the discovery limits for the 
transverse momentum variables $p_T^\gamma$, $p_T^\ell$ and $\not{\!p}_T$ in the two 
cases ($a$) $\Delta\kappa_\gamma \neq 0, \lambda_\gamma = 0$ and 
($c$) $\Delta\kappa_\gamma = 0, \lambda_\gamma \neq 0$ respectively. 
Corresponding limits for the azimuthal angle variables $\Delta \varphi_{\gamma\ell}$, 
$\Delta \varphi_{\gamma\;\not{\!p}_T}$ and $\Delta \varphi_{\ell\;\not{\!p}_T}$
are similarly shown in the lower panels, marked ($b$) and ($d$) respectively.   
As before, the CMS limits from $W\gamma$ production\cite{CMS-WA}, as well as those from
$WW$ production\cite{CMS-WW} are shown by solid and broken vertical lines respectively.
As in Figure~\ref{fig:total-cs}, a broken horizontal line represents the maximum
integrated luminosity envisaged at the HL-LHC, and therefore, its intersections
with the different curves indicates the discovery limit of the machine. 

If we now inspect the discovery limits in Figure~\ref{fig:limits-single} and 
compare then with those in Figure~\ref{fig:total-cs}, the following conclusions
emerge.
\vspace*{-0.2in}
\begin{itemize}
\item For $\Delta\kappa_\gamma < 0, \lambda_\gamma = 0$, the discovery
limits from the total cross-section are better than those from the 
distributions; among the distributions, the best constraints arise
from $\Delta \varphi_{\ell\;\not{\!p}_T}$. 
\item For $\Delta\kappa_\gamma > 0, \lambda_\gamma = 0$, the discovery
limits from the total cross-section are no longer better; instead the
$p_T$ distributions are more efficient, especially as the luminosity
increases above 100~fb$^{-1}$. The $\Delta \varphi_{\ell\;\not{\!p}_T}$
distribution can be used to get discovery limits comparable to those
from the different $p_T$ distributions, but not better.    
\item For $\Delta\kappa_\gamma = 0, \lambda_\gamma < 0$, the total
cross-section and the $p_T$ distributions give similar discovery limits,
while the discovery limits from the $\Delta \varphi_{\ell\;\not{\!p}_T}$
are significantly better and obviously improve as the integrated
luminosity increases.
\item For $\Delta\kappa_\gamma = 0, \lambda_\gamma > 0$, the total
cross-section gives better discovery limits than the $p_T$ distributions,
whereas the $\Delta \varphi_{\ell\;\not{\!p}_T}$ distribution always gives
better sensitivity.  
\end{itemize}
\vspace*{-0.2in}
It is also interesting to note that of the three $p_T$ distributions, the
best results are obtained from different distributions in different
regimes, whereas for the $\Delta\varphi$ distributions,  
$\Delta \varphi_{\ell\;\not{\!p}_T}$ is always the most sensitive. 
This sensitivity is likely to be due to interference between different helicity
amplitudes \cite{Panico}, though that is not explicit in our calculations.

There is a very important lesson to learn from the above observations,
viz., that there is no unique variable whose study will provide the 
maximum sensitivity to anomalous TGCs. A proper experimental study
should, therefore, include {\it all} the variables considered above, including
the total cross-section. Currently, experimental results are mostly
based on transverse momentum studies\cite{LEP-2,ATLAS-WA,CMS-WA}, but 
these, as our results indicate, are not always the most sensitive variables.

Thus far, we have only considered one of the TGCs at a time, viz., either
$\Delta\kappa_\gamma \neq 0, \lambda_\gamma = 0$, or $\Delta\kappa_\gamma = 0, 
\lambda_\gamma \neq 0$. While convenient from a purely phenomenological
standpoint, this is hard to justify from a top-down approach, for the
same new physics which creates nonzero $\Delta\kappa_\gamma$ could very
well generate nonzero $\lambda_\gamma$ as well. We now turn, therefore, to 
the study of this more realistic case of joint variation of the two parameters.
The formulae in Eqns.~(\ref{eqn:chi2tot}) and (\ref{eqn:chisqdistrib}) are
naturally geared to handle this joint variation, so all that is required
is to numerically vary both the parameters and perform the same kind of analysis as
we have described above.

\begin{center}
\begin{figure}[!htb]
\centerline{\includegraphics[width=0.95\textwidth]{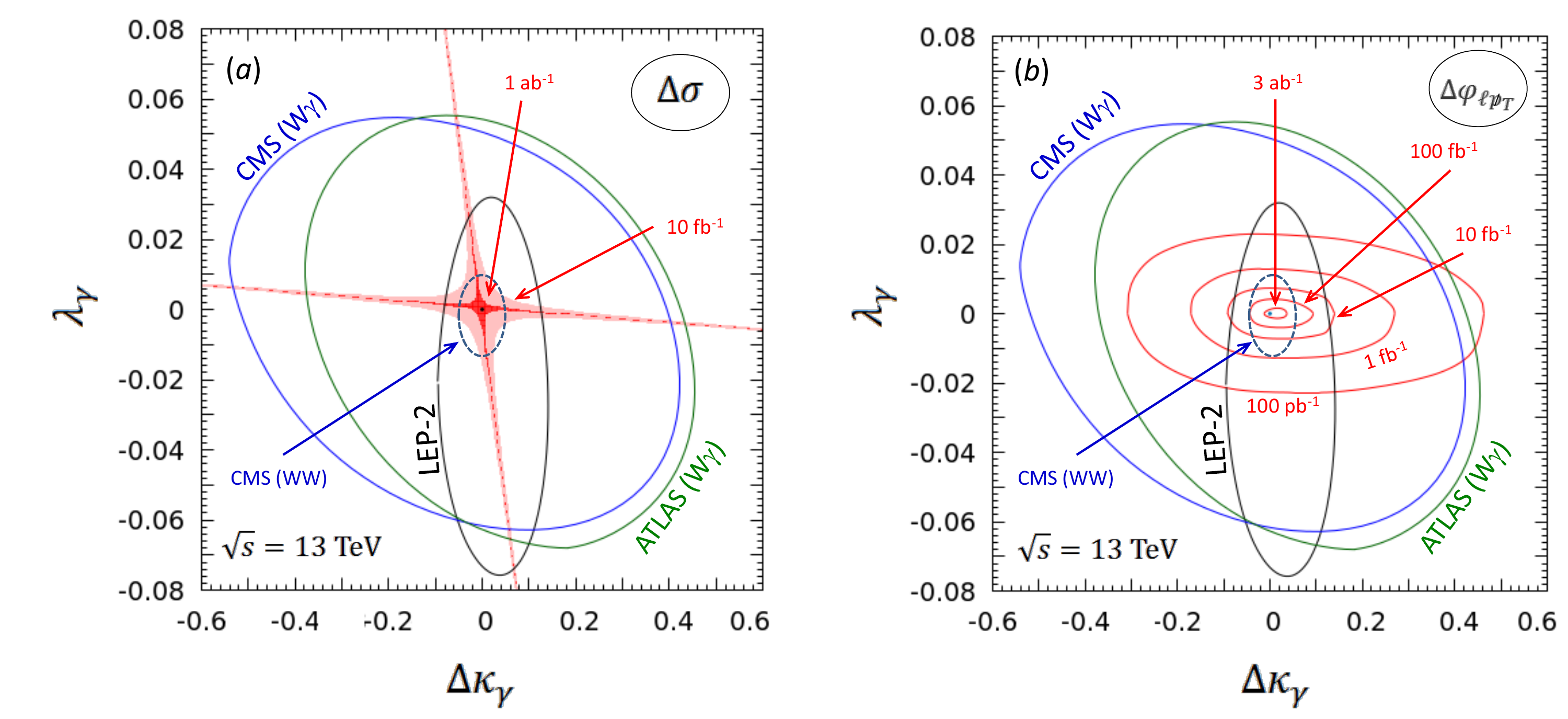} }
\vspace*{-0.1in}
\caption{\setstretch{1.05}\footnotesize Joint discovery limits at 95\% C.L. on the
anomalous couplings $\Delta\kappa_\gamma$ and $\lambda_\gamma$. The measurables used
are ($a$) the total cross-section, and ($b$) the azimuthal angle variable
$\Delta \varphi_{\ell\not{\!p}_T}$ respectively. In the left panel, marked ($a$)
the regions shaded pink (red) are inaccessible to the LHC with 10 (1000) fb$^{-1}$ of 
integrated luminosity. Similar inaccessible regions lie inside the oblate ellipses 
(in red) on the right panel, marked ($b$). Experimental constraints from LEP-2
and from the Run-1 of LHC are shown as prolate ellipses in both panels. The tiny
black dot at the centre is, of course, the SM prediction at tree-level. } 
\label{fig:limits-double}
\end{figure}
\end{center}

Our results for joint variation are shown in Figure~\ref{fig:limits-double}.
The left panel, marked ($a$) shows the discovery limits that can be obtained 
using the total cross-section. The inaccessible region at the 13~TeV LHC,
assuming an integrated luminosity of 10(1000)~fb$^{-1}$ is shaded in
pink(red). For comparison, on the same panel 
we give the constraints from LEP-2 (black), and
from the CMS (blue) and ATLAS (green) Collaborations at the LHC Run-1. In each
case the inside of the ellipse is not accessible and the region outside is ruled 
out. It is immediately obvious that, as was the case with one parameter at a time,
the total cross-section is a reasonably sensitive probe of anomalous TGCs,
and in fact, even with 10~fb$^{-1}$ of data, it is as sensitive as the use of the
$WW$ production data (modulo the $WWZ$ caveat). Sensitivity improves dramatically
for 1000~fb$^{-1}$ luminosity, as the tiny red shaded region indicates. However
-- and here lies the rub -- the inaccessible region is star-shaped, with four
arms which stretch to possible large values of one of the parameters at a time. 
It is easy to see why these arise, for the significance is based on a single parameter, 
viz., the total cross-section, and there will always be regions where the contributions
to this from $\Delta\kappa_\gamma$ cancel with those from $\lambda_\gamma$, making
the signal small or vanishing. Thus, although the total cross-section can be used
to probe the anomalous TGCs quite efficiently, there remain these four narrow wedges
of the parameter space which are inaccessible to the LHC.    

\vspace*{-0.6in}

The situation can be radically improved by using a distribution, rather
than the total cross-section, for it is almost inconceivable that the extra
contributions from $\Delta\kappa_\gamma$ will undergo a bin-by-bin cancellation with 
those from $\lambda_\gamma$, given that the distributions are somewhat different, 
as shown in Figures~\ref{fig:distrib-kap} and \ref{fig:distrib-lam}. To be precise,
the same pair of values which cause cancellation of anomalous effects in one
bin, may not cause cancellation in another bin, and hence, the overall value of 
$\chi^2$ will not be rendered small. This is illustrated in the right panel, marked ($b$)
of Figure~\ref{fig:limits-double}, where we use the distribution in  
$\Delta \varphi_{\ell\;\not{\!p}_T}$ to obtain 95\% C.L. discovery limits. Here,
corresponding to different values of the integrated luminosity, we show the
discovery limits as elliptic regions in the same way as shown by the experimental
collaborations. As usual, the interior of each ellipse is inaccessible to the
LHC with the luminosity in question. The experimental constraints are given
exactly as in the left panel, marked ($a$). It hardly needs to be commented that at the
HL-LHC, very stringent constraints indeed could be obtained in case no deviation
from the SM is seen. It may be noted, however, that even with this accuracy of
measurement, the one-loop SM effects will not be accessible, though 
effects from new physics such as the MSSM, may be\cite{LEP-Kneur} .       

To summarise, then, we have considered the process $pp \to \gamma W^\ast \to 
\gamma \; \ell 
\!\not{\!p}_T$ at the 13~TeV run of the LHC, and studied possible implications of having
anomalous ($CP$-conserving) $WW\gamma$ vertices in the theory. The choice of this
process (which has a lower cross-section than, say, $W^+W^-$ pair production) is 
because the tagging of a final-state photon ensures that there is no contamination
of the new physics contribution with possible anomalous effects in the $WWZ$ vertex.
We have shown that the anomalous $WW\gamma$ couplings may be constrained by considering 
not one,
but seven independent observables, viz. the total cross-section, three different $p_T$
distributions and three different azimuthal angle variables. The relative efficacy of
each of these has been studied in detail, making certain simplifying assumptions, such
as the absence of initial/final state radiation, pileup effects, systematic errors
and detector effects. The first two we expect to be essentially eliminated by the 
rather severe kinematics cuts chosen for our analysis, but the latter ones can only
be estimated by a thorough experimental analysis, which is beyond the scope of this
work. Similarly, we have assumed that the kinematic cuts suggested by us will be 
effective in controlling backgrounds from $W$ + jet events (with a jet faking a 
photon). Under these assumptions, we have shown that the judicious use of the 
variables studied, especially the azimuthal angle variable 
$\Delta \varphi_{\ell\;\not{\!p}_T}$, can be used to pinpoint anomalous effects
in the process in question, to a great degree of accuracy, as the statistics 
collected by the LHC (and its HL upgrade) grow larger. Such measurements would
eventually probe not just large electoweak corrections in the TGC sector, but
could also effectively constrain new physics involving  modifications and mixings in 
the gauge sector. Of course, the most exciting scenario would be to see an unambiguous
deviation from the SM prediction in any of the variables (or more than one variable)
in the upcoming runs of the LHC, and it is on this hopeful note that we conclude this
article.   

\bigskip
 
\noindent{\small {\sl Acknowledgements}: The authors are grateful to Sandhya Jain and 
Gobinda Majumder (CMS Collaboration) for discussions.}

\end{document}